\documentclass[manuscript]{acmart}
\usepackage{soul,xcolor}

\renewcommand\footnotetextcopyrightpermission[1]{} 
\usepackage{tikz}
\newcommand*\circled[1]{\tikz[baseline=(char.base)]{
            \node[shape=circle,fill,inner sep=0.5pt] (char) {\textcolor{white}{#1}};}}
\newcommand*\circledwbg[1]{\tikz[baseline=(char.base)]{
            \node[shape=circle,fill=white,draw=black,inner sep=0.5pt] (char) {\textcolor{black}{#1}};}}

\begin{document}

\title{An Inference and Learning Engine for Spiking Neural Networks in Computational RAM (CRAM)}

%
\author{H\"{u}srev C{\i}lasun}
\email{cilas001@umn.edu}
\author{Salonik Resch}
\email{resc0059@umn.edu}
\author{Zamshed Iqbal Chowdhury}
\email{chowh005@umn.edu}
\author{Erin Olson}
\email{olso6834@umn.edu}
\author{Masoud Zabihi}
\email{zabih003@umn.edu}
\author{Zhengyang Zhao}
\email{zhaox526@umn.edu}
\author{Thomas Peterson}
\email{pete9290@umn.edu}
\author{Keshab Parhi}
\email{parhi@umn.edu}
\author{Jian-Ping Wang}
\email{jpwang@umn.edu}
\author{Sachin S. Sapatnekar}
\email{sachin@umn.edu}
\author{Ulya Karpuzcu}
\email{ukarpuzc@umn.edu}
\affiliation{%
  \institution{University of Minnesota}
  \city{Twin Cities}
  \state{Minnesota}}

\renewcommand{\shortauthors}{C{\i}lasun, et al.}

\begin{abstract}
Spiking Neural Networks (SNN) represent a biologically inspired computation model capable of 
emulating neural computation in human brain and  brain-like structures. The main promise is very low energy consumption. Unfortunately, classic Von Neumann architecture based SNN accelerators often fail to address demanding computation and data transfer requirements efficiently at scale. In this work, we propose a promising alternative, an in-memory SNN accelerator based on Spintronic Computational RAM (CRAM) to overcome scalability limitations, which can reduce the energy consumption by up to {164.1}$\times$ when compared to a representative ASIC solution.
\end{abstract}




\maketitle

\section{Introduction}

Spiking Neural Networks (SNN) are neural networks mimicking the impulse based neural transmission in the brain. Recently, as a computational model, SNNs have been emerged as biologically more realistic, yet tractable alternatives to artificial neural networks. Although there are topological similarities to a fully-connected neural network, the main difference lies in the \emph{spike} event, an electrical discharge triggered by a series of chemical reactions. Spike trains coming from \emph{presynaptic neuron}s are processed in the \emph{postsynaptic neuron}, which enables 
computational tasks 
such as classification to be performed in SNN. 

As a biologically inspired computational model capable of emulating neural computation in human brain and brain-like structures \cite{furber2014spinnaker,stewart2012spaun}, 
inherent SNN architecture poses as an ideal computing medium for event-driven processing~\cite{stromatias2017event}. As a result,
SNN 
can perform 
a variety of computational tasks with significantly lower power consumption, such as prosthetic brain-machine interface control \cite{dethier2011brain} and speech recognition \cite{yin2017low}. 
Recent efforts \cite{lee2018flexon} 
focus on flexible and efficient hardware emulation of different biologically accurate SNN models, while others 
exploit the computational power resulting from the high number 
of relatively simpler neurons in SNNs. 
Low energy hardware SNN accelerators such as
ODIN \cite{frenkel20180}
try to minimize the energy per spiking operation.
SNN hardware solutions
include IBM's TrueNorth \cite{merolla2014million}, Intel's Loihi \cite{davies2018loihi}, University of Manchester's SpiNNaker \cite{furber2014spinnaker}, and {Human Brain Project's BrainScaleS} \cite{brainscales}.
Present highest number of spiking neurons in a hardware SNN implementation is 1 Billion {real-time neurons in 
{SpiNNaker}} \cite{jin2010modeling},
only a fraction of the average neuron count of 86 Billion in human brain \cite{lent2012many}. 

Large-scale SNNs needed to perform useful computational tasks
{inevitably come with}
increased data access and parallelism demand which
{often exceeds capabilities of}
traditional hardware.  When neuron count increases significantly, {more space is required to store} synaptic parameters.
Data access {and retrieval} becomes a burden at the same time,
{mainly because any spiking information emerging in neurons should be accessible to all other neurons as an input.}
Various solutions based on traditional CMOS as well as emerging spintronic and resistive technologies are proposed to overcome this bottleneck \cite{BenchmarkingNeuralInference}. 
For example, recent
{Magnetic Tunneling Junction (MTJ) based spintronic}
SNN accelerators
\cite{sengupta2016hybrid,chen2018magnetic,yang2020all} 
multiply weights 
by spike trains using MTJs in a crossbar setting.
These and similar designs 
typically only address a single stage of SNN computation -- i.e., 
weight multiplication by spike train.
Flexibility
is also a concern,
{as the SNN model as well as parameter resolution is 
hardwired.}
Spintronic 
{(MTJ based) devices} have also been proposed as analog SNN building block{s} \cite{zhang2016stochastic}, however, analog implementations by construction suffer from process variability constraints \cite{snnstt}. 

Putting it all together, large scale SNN computations are massively parallel and induce high-intensity memory accesses for data retrieval. As a result, the energy consumption skyrockets at scale.
This is why true in-memory computing substrates such as the recently proposed spintronic Computational RAM (CRAM) \cite{wang2015general}, which enable massively parallel and reconfigurable logic operations {\em in-situ} without compromising energy efficiency, represent especially promising platforms for SNN hardware, which form the focus of this paper.   
We will start our discussion with background information covering basics of CRAM and the SNN model in
Section~\ref{sec:background}.
Section \ref{sec:cramsnn} details the proposed SNN architecture based on CRAM. After quantitative evalution in 
Section \ref{sec:eval}, we conclude the paper in  
Section \ref{sec:conclusion}.


\section{Background} \label{sec:background}

After covering CRAM basics, we will continue with 
the feedforward Leaky Integrate-and-Fire SNN model in Section \ref{sec:background:liafsnn}. Although only the feedforward model is useful for many computational tasks, applications such as brain simulation needs the parameters to be learned online. To this end, we also include pairwise Spike-Time Dependent Plasticity (STDP) learning algorithm for updating parameters during computation, as need be.

\subsection{CRAM Basics} \label{sec:background:cram}

Essentially, CRAM \cite{wang2015general} augments conventional spintronic memory arrays with compute capability, thereby enabling seamless memory access.
As long as there is no computation, CRAM reduces to an ordinary memory. When computation is enabled, CRAM performs logic operations directly inside the memory array.
Spin Torque Transfer (STT) and Spin-Hall Effect (SHE) or Spin-Orbit Torque (SOT) based CRAM variants exists \cite{chowdhury2019jxcdc,ReschPIMBALL,zabihi2019,zabihiSHE}. 
CRAM can perform one logic operation (Boolean gate) in a column at a time, but all (or a desired subset of) columns can perform the same operation in parallel. Each cell in a column can serve as an input or output to a Boolean gate. Logic operations are reconfigurable in time and space. At the same time, multiple CRAM arrays can perform the same computation (across all of their columns) in parallel.

CRAM's main storage element is a Magnetic Tunneling Junction (MTJ), which comprises a fixed-polarity magnetic layer, a variable-polarity magnetic layer, and an insulating layer in between.
When polarities of the magnetic layers (mis)match, the MTJ is in (Anti)Parallel-(A)P  state which corresponds to logic (1) 0, exhibiting a (high) low resistance. 

\begin{figure}
    \centering
    \includegraphics[trim={0.82cm 23cm 12.3cm 0.7cm},clip,width=8cm]{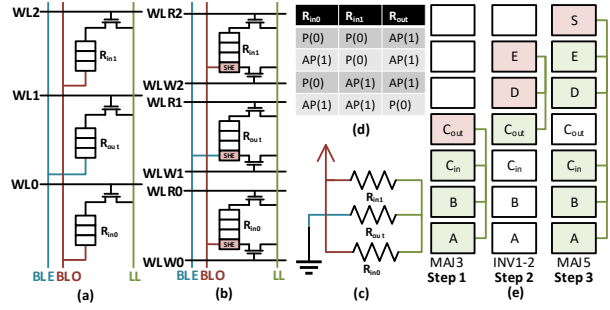}
    \caption{(a) \texttt{1T1M} STT-CRAM;
    (b) SHE-CRAM; (c) Equivalent resistive network for a \texttt{NAND} gate; 
    (d) \texttt{NAND} truth table;
    (e) 3-step full adder implementation.
    {\texttt{MAJ3} and \texttt{MAJ5} are 3-input and 5-input majority gates. 
    \texttt{INV1-2} is 
    an inverter which writes the output to two different cells simultaneously.}}
    \label{fig:cram}
\end{figure}

Cell structures for STT- and SHE-CRAM are provided in Figure \ref{fig:cram}(a) and \ref{fig:cram}(b), respectively.
{STT-CRAM features even and odd bitlines (BLE/O), which are used to sense (change) the state of the MTJ in read (write) mode. For logic operations, BLE/O is used determine which MTJ serves as an input or output. Logic lines (LL) 
connect input and output cells to perform Boolean operations. 
STT-CRAM also has wordlines (WL), which are used to select the rows for both memory and logic operations.} 

Specifically, in order to perform a logic gate in Figure \ref{fig:cram}(a), first the output MTJ is  preset to a known logic value
depending on the type of the gate.
If inputs reside in even rows, the output should be in an odd row, and vice versa. 
{By imposing a specific voltage difference between BLE and BLO, a current is induced through (a parallel connection of) inputs and the output in series. The magnitude and direction of this current depends on the voltage difference between BLE and BLO, which also is a function of the type of the gate to be performed. For a given voltage difference, the current through the output evolves as a function of the parallel equivalent resistance of the inputs, and may or may not be enough to switch the output state. This is the main principle behind how CRAM implements truth tables. 
 If the combined current is high enough to change the state of the output cell(s), the cell state is changed,
effectively corresponding to the implementation of a logic gate. 
}
The equivalent
resistive network 
for the universal NAND gate 
is given in Figure \ref{fig:cram}(c) along with  the truth table in Figure \ref{fig:cram}(d).
CRAM supports a rich set of (universal) gates beyond NAND, each characterized by specific bitline voltage and output preset values. More complex logic functions such as full-adders can be synthesized by a sequence of basic gate operations as shown in Figure \ref{fig:cram}(e) \cite{ChowdhuryCompArchLett}.

SHE-CRAM, as shown in Figure \ref{fig:cram}(b), on the other hand, features a four terminal 
cell element which enables
faster memory and logic operations while consuming significantly less energy. 
This is because SHE-CRAM separates read and write paths, making independent optimization of each possible (which otherwise impose conflicting constraints).
Accordingly, SHE-CRAM has separate read (WLR) and write (WLW) wordlines, and two access transistors per cell. SHE-CRAM hence has a higher area footprint compared to the STT-CRAM, which is compensated by lower energy and faster operation. Memory and logic operations follow the same principles otherwise. 

To ease illustration, in the following, we will use \emph{transposed} CRAM arrays, and abstract out the {interleaved} placement where inputs (outputs) reside in even (odd) rows or vice versa.

%

\subsection{Leaky Integrate-and-Fire SNN Model} \label{sec:background:liafsnn}
Although there exist more complex and biologically accurate models for SNNs, {\em Leaky Integrate-and-Fire} model, as described in \cite{davies2018loihi}, is widely used due to its relative simplicity. The general model assumes that each neuron is connected to \emph{presynaptic neuron}s which broadcast their own spike trains. Each connection from  neuron $j$ to neuron $i$ is characterized {by} a weight $\omega_{ij}$ and a delay value $d_{ij}$ corresponding to the actual transmission delay in brain for more accurate modeling. Each spike {train} (or spike function)
can be expressed as $\sigma(t)=\sum_k{\delta(t-t_k)}$ where $\delta$ is the unit impulse function; $t$, the discrete time variable; and $t_k$, the time difference corresponding to $k^{th}$ spike. Table~\ref{tab:params_defs} provides the definition for all model parameters. 

\begin{table}[ht]
\caption{Parameter Definitions}
    \centering
    \scalebox{0.95}
    {
    \begin{tabular}{c|c}
         Parameter & Definition  \\
         \hline
         $t$ & discretized time variable\\
         $\sigma_j(t)$ & spike train at the output of neuron $j$\\
         $\tau_u$ & neural time constant\\
         $\tau_v$ & synaptic time constant\\
         $\theta_i$ & spiking threshold constant of neuron $i$\\
         $\omega_{ij}$ & weight variable between neurons $i$ and $j$\\
         $d_{ij}$ & delay variable between neurons $i$ and $j$\\
         $b_{i}$ & bias variable of neuron $i$\\
         $u_i(t)$ & synaptic response current of neuron $i$\\
         $v_i(t)$ & membrane potential of neuron $i$\\
         $A_{\pm}$ & time constant\\
         $t^{pre/post}$ & presynaptic/postsynaptic spike time\\
         $\alpha_u(t)$ & $\frac{1}{\tau_u} e^{-\frac{t}{\tau_u}} H(t)$\\
         $\mathcal{F}(t)$ & $e^{-\frac{t}{\tau_u}}H(t)$\\
         $L_f$ & \#entries in the lookup table of $\alpha_u(t)$\\
         $S$ & weight bit size\\
         $t_{max}$ & maximum time difference for STDP\\
         $r(t)$ & pseudorandom noise
    \end{tabular}
    }
    \label{tab:params_defs}
\end{table}

{The weighted sum of filtered presynaptic spike trains with bias is called \emph{synaptic response current}}, as depicted in  
Equation \eqref{eq1},
where
$u_i(t)$ represents the synaptic response current of neuron $i$:
\begin{equation} \label{eq1}
u_i(t) = \sum_{j\neq i} \omega_{ij}(\alpha_u \ast \sigma_j )(t)+b_i
\end{equation}
$\alpha_u(t)=\frac{1}{\tau_u} e^{-\frac{t}{\tau_u}}H(t)$
where $H(t)$ is the unit step function;
$\tau_u$, a time constant; and $b_i$, the bias.

The voltage difference between the inside and the outside of the synapse is called \emph{membrane potential}.
Update function for membrane potential is given as -- where $v_i(t)$ captures the  membrane potential of neuron $i$:
\begin{equation} \label{eq2}
v_i(t) = -\frac{1}{\tau_v}v_i(t)+u_i(t)-\theta_i \sigma_i(t) 
\end{equation}
$\tau_v$ here is a time constant; $\theta_i$, the threshold value for neuron $i$. If the neuron spikes, $v_i$ is initialized to zero. 

\begin{figure*}[htp!]
    \centering
    \includegraphics[trim={0cm 11.2cm 49.9cm 0cm},clip,width=13.9cm]{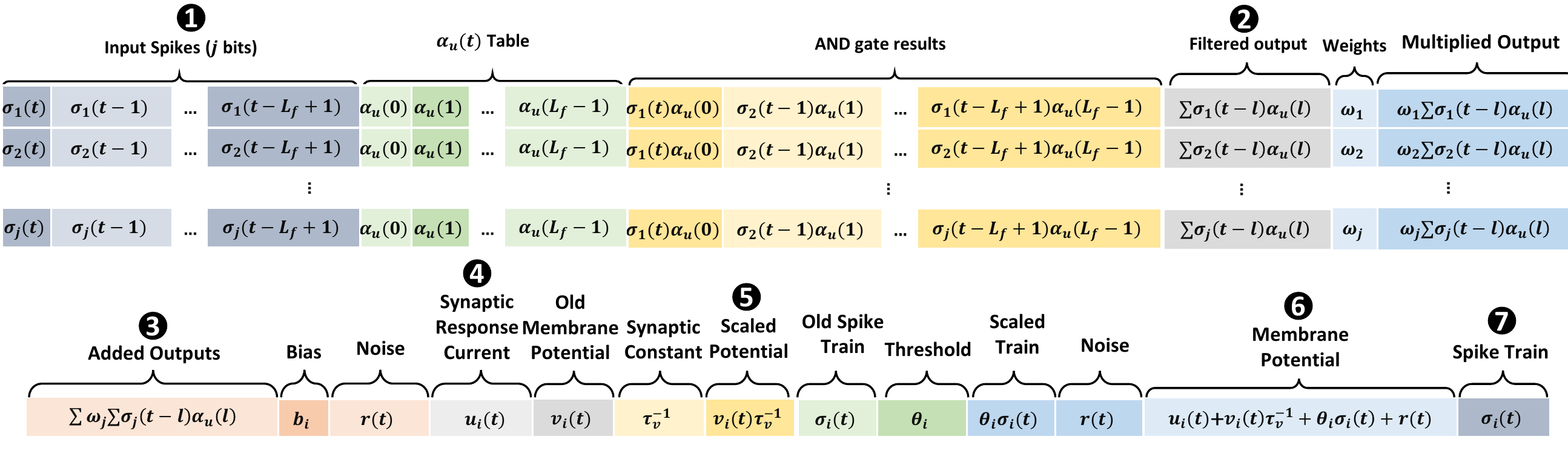}
    \caption{Data layout (transposed).}
    \label{fig:layout}
\end{figure*}

\begin{figure}[htp!]
    \centering
    \includegraphics[trim={0cm 9.6cm 50.8cm 0cm},clip,width=8.1cm]{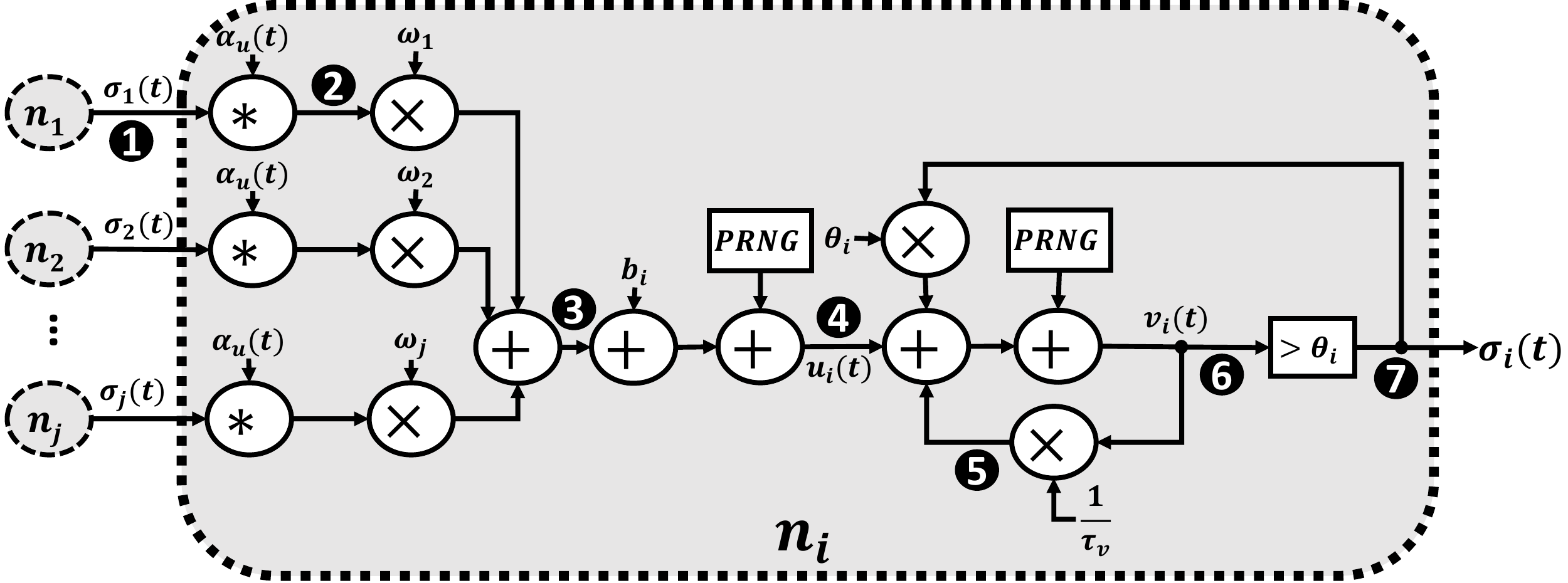}
    \caption{Leaky Integrate-and-Fire model \cite{davies2018loihi}. 
    }
    \label{fig:scheme}
\end{figure}

Figure {\ref{fig:scheme}} depicts an overview of the model. Putting it all together, each neuron calculates its output spike train from presynaptic spike trains.
Variations of this  basic algorithm include the Random Sampling algorithm from Loihi \cite{davies2018loihi}, which our design is based on, and where
synaptic response current and membrane potential are optionally incremented by
a pseudorandom number 
-- which we will encapsulate in the following discussion in a noise term $r(t)$. 

\subsection{Pairwise STDP based Parameter Learning}

%
 {Learning entails adjusting synaptic weights ($\omega_{ij}$) and delays ($d_{ij}$) to dynamically optimize the neural network for solving a specific problem. It models the actual synaptic changes which brain adapts to accommodate new constraints.} 

{STDP} model features a simple weight learning rule, as described in \cite{davies2018loihi,dang2020efficient}. It can be summarized as follows
\begin{equation} \label{eq:stdp}
 \Delta \omega_{i,j} = \begin{cases} 
      A_{-}\mathcal{F}(t-t_i^{post}), & \text{on presynaptic spike} \\
      A_{+}\mathcal{F}(t-t_j^{pre}),  &  \text{on postsynaptic spike}
   \end{cases}
\end{equation}
where $\mathcal{F}(t)=e^{-\frac{t}{\tau}}H(t)$ and $\tau$, $A_{-}$, and $A_{+}$ are constants, as defined in Table~\ref{tab:params_defs}. $\Delta \omega_{i,j}$ is added to the weight in each update. 

Figure \ref{fig:stdp_alternative} provides a
block diagram description of the basic STDP algorithm.
{
{We define $\Delta t^{pre(post)}$ as $t-t_{j(i)}^{pre(post)}$.}
The algorithm continuously checks
whether the current neuron or a presynaptic neuron spikes. If so, it resets the $\Delta t^{pre(post)}$ counter by multiplying its former value by zero. Otherwise, it checks for overflow and increments the $\Delta t^{pre(post)}$, which is then fed to $\mathcal{F(\cdot)}$ function and multiplied by $A_{+(-)}$.}

\section{SNN in CRAM} \label{sec:cramsnn}

\subsection{Leaky Integrate-and-Fire-Model in Memory} \label{sec:cramsnn:iafinmem}


CRAM supports universal Boolean gates, hence can handle any type of computation including 
 the Leaky Integrate-and-Fire Model described in Section \ref{sec:background:liafsnn}. The mapping is straightforward, as shown in Figure \ref{fig:layout}, where
  each neuron gets 
 processed inside a single CRAM array to exploit array-level parallelism. 
 
 \begin{table}[ht]
    \centering
    \caption{{CRAM array utilization by constants, parameters, and lookup table for different $S$ and $L_f$s.}}
    \scalebox{0.97}{
    \begin{tabular}{c|c|c|c}
        $S$ & $L_f$ & Array Size & Utilization \\
         \hline
         $1-4$ & $32$ & $256\times256$ & $14.06\%-56.25\%$\\ 
         $5-8$ & $32$ & $512\times512$ & $35.15\%-56.25\%$\\ 
         $9-16$ & $32$ & $1024\times1024$ & $31.64\%-56.25\%$\\ 
         $1-2$ & $64$ & $256\times256$ & $26.55\%-53.13\%$\\ 
         $3-4$ & $64$ & $512\times512$ & $39.84\%-53.13\%$\\ 
         $5-8$ & $64$ & $1024\times1024$ & $33.2\%-53.13\%$\\
         $9-16$ & $64$ & $2048\times2048$ & $29.88\%-49.8\%$\\
    \end{tabular}
    }
    \label{tab:utilization}
\end{table}
 
\subsubsection{Initialization}
\label{sec:cramsnn:ffinit}
Our design incorporates several lookup tables (LUT) and dedicated storage for parameters and constants, which are initialized before feedforward computations start. 
The initialization is a one-time procedure, as the corresponding values do not change during computation. 
{
We also initialize an $S$-bit \emph{local delay} value for each synapse, which serves modeling synaptic delays between neurons.}

Precomputed $\alpha_u$ values reside in a lookup table. $\alpha_u$ lookup table consists of $L_f$ different $S$-bit values (Table~\ref{tab:params_defs}). Similarly, constant $\frac{1}{\tau_v}$ is initialized once, as well as weights $\omega_{ij}$, $b_i$, and $\theta_i$ values. 
For the $\alpha_u$ lookup table, Table \ref{tab:utilization} shows the utilization for different CRAM array sizes for various $S$ and $L_f$. We 
choose the array size so that the utilization does not exceed approximately half of the array size. Thereby enough space is left to perform the arithmetic/logic operations in remaining CRAM space.

\subsubsection{Data Flow}
Once a spike train of presynaptic neurons is received, it is written in the memory as shown in the first column 
in Step \circled{1}. 
Along with all the previously received (and similarly placed) spikes,
these spikes and the $\alpha_u(s)$ values are multiplied where $s \in [0,L_f-1]$ and $L_f$ is the predetermined filter length. 
{This filtering operation effectively corresponds to the convolution ($*$) from Equation \eqref{eq1}.} Since all of the values involved are binary, this operation reduces to $S\times L_f$ \texttt{AND} operations where $S$ is the bit length of each entry in the $\alpha_u(s)$ lookup table, which keeps
precomputed values of the $\alpha_u(s)$ function.
Once \texttt{AND}s are computed, the resulting $L_f$ entries are added together and {rounded} (Section~\ref{sec:lif_lowlevel}) 
to $S$ bits to obtain the column denoted by \circled{2} in Figure \ref{fig:layout}. In order to prepare the spike
train for the next spike {computation cycle}, the spike train is next
shifted by reading the columns obtained
in Step \circled{1} and writing them to 
{another row}. Then, the resulting $S$ bit values are multiplied by $S$ bit weights, which corresponds to $S^2$ full adder operations and results in the column marked by Step \circled{3}. The $2S$ multiplication outcome is next rounded to $S$ bits, i.e., a $(S-1)$-bit rounding factor is added to it and the result is truncated to $S$ bits. Addition is again performed as a cascade of full adder operations. Truncation has practically no overhead as it translates into simply ignoring the unused bits.
 Equation \eqref{eq1} gives rise to the operations we covered so far, spanning Step \circled{1} to Step \circled{3}.
 
 In the following step, all rows 
 in Step \circled{3} are added by reading half of the rows and writing them back to the adjacent rows. After each addition, a rounding operation is performed. This operation takes $\log_2 j$ stages where $j$ is the predetermined maximum number of presynaptic neurons. Once the addition is done, the result is reduced to a single row as shown in Step \circled{4}. After the bias and the {noise} is added to the outcome from Step \circled{4}, synaptic response current shown in Step \circled{5} is obtained. 
 We keep older membrane potential in the same row, which is next 
 scaled -- by 
 multiplication with ${\tau_v}^{-1}$. Previous spike train
 is multiplied by the threshold value $\theta$, which reduces to an \texttt{AND} operation of $S$ bits. 
 Equation \eqref{eq2} spans Step \circled{4}--Step \circled{6}.
 The current membrane potential is obtained after current synaptic response current, scaled old membrane potential, the noise, and the scaled old spike train is added as shown in Step \circled{6}. Current membrane potential then  overwrites the old membrane potential. Finally, current spike value is calculated by thresholding the current membrane potential in Step \circled{6}, which corresponds to the comparison of $S$-bits. Current spike value is written as shown in Step \circled{7} and copied to the old spike train column. 
 {In order to reset the membrane potential, we invert the spike bit and \texttt{AND} with the membrane potential.} The output spike is then read and broadcasted as we will explain in Section \ref{sec:cramsnn:routing}.
 
 {Before starting computation,   the $S$-bit local delay value in all columns is incremented and compared to $d_{i,j}$ (for the corresponding synaptic connection).
 If the comparison yields true, the local delay value is reset to zero. The comparison output is then read by the array controller and used as column enable. This results in a practical synaptic delay implementation, as well as energy savings, as the disabled columns do not participate in computations described above to perform Equations \eqref{eq1} and \eqref{eq2}.}

\subsubsection{Low Level Operations} \label{sec:lif_lowlevel}
{For addition operations, we use the full adder from Figure \ref{fig:cram}. Multiplication also uses the same full adder design $N^2$ times for $N$-bit numbers. In the convolution with $\alpha_u(t)$, the multiplications reduce to \texttt{AND} operations since the spikes are binary. Comparison by the threshold value is implemented as a cascade of $6N-3$ \texttt{NAND} gates for $N$-bit numbers. We increase the bit size conservatively until Step \circled{3} and then we perform rounding operation by adding rounding factors in each arithmetic operation, which ensures that no overflow happens.} 
{Rounding to $S$ bits entails adding a $(S-1)$ bit long rounding factor and then keeping only the $S$ most significant bits.}


\begin{figure}[ht]
    \centering
    \includegraphics[trim={0cm 6cm 9.9cm 0cm},clip,width=8.5cm]{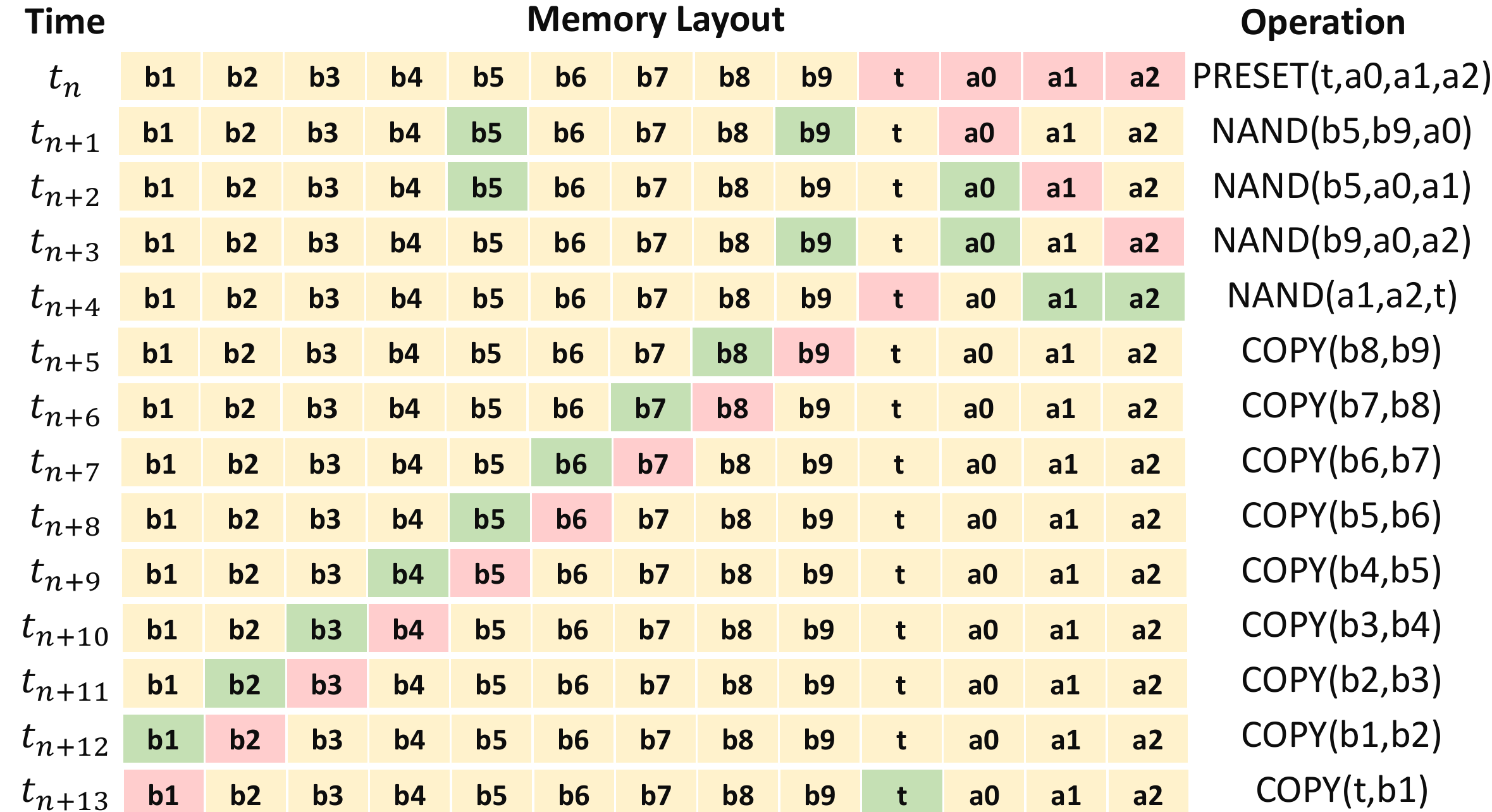}
    \caption{Linear Feedback Shift Register (LFSR) based pseudorandom number generation in CRAM using feedback polynomial $x^9+x^5+1$. Input (output) cells are highlighted with green (red).}
    \label{fig:prng}
\end{figure}

{Loihi \cite{lines2018loihi} features a Pseudorandom Noise Generator (PRNG) to support several variants of the basic algorithm such as neural sampling. As a proof of concept, we demonstrate how to implement a PRNG in CRAM}
using a basic linear feedback shift register 
in Figure \ref{fig:prng}. First, an XOR gate is applied as a cascade of four \texttt{NAND} gates and then \texttt{COPY} operations are performed as many as the length of the feedback polynomial. For an example 9-bit polynomial, this operation takes 13 cycles. {We denote this generated noise with $r(t)$ and update each time it is used.}

{Our lookup table based implementation of $\alpha_u(t)$ function and the inevitably limited bit length for parameters such as weights, by construction, affect
synaptic accuracy. Section \ref{sec:eval} provides a quantitative characterization of the accuracy impact 
of precision.
}

\begin{figure}[ht]
    \centering
    \includegraphics[trim={0cm 8cm 19cm 0cm},clip,width=8.3cm]{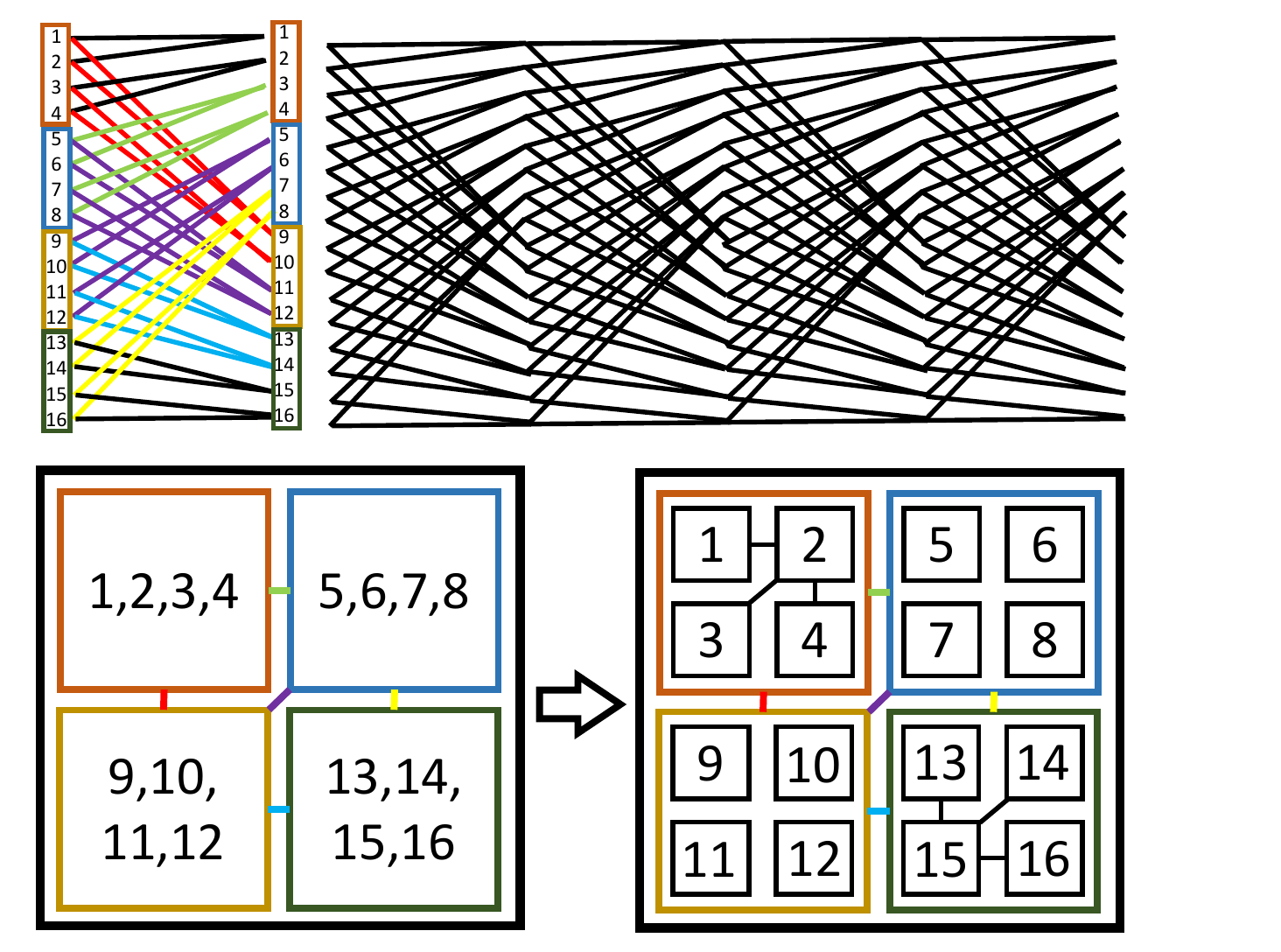}
    \caption{Generalized De Bruijn Graph for an example 16-array network, self connections are not shown.}
    \label{fig:gdbg2}
\end{figure}

\subsection{Routing and Connectivity} \label{sec:cramsnn:routing}
 As one CRAM array is allocated to process each neuron,
 {the basic connectivity between neurons directly translates into the connections between arrays to transfer spikes.} 
 Ideally, each neuron in SNN
 would be connected to all of the other neurons. However, such a fully connected scheme requires 
 {$\binom{N}{2}$} connections between neurons, which is not feasible in hardware when the number of neurons is 
 very high (i.e., in the order of Billions), 
 {due to the limited area.}
 In order to eliminate the connectivity burden,
 a possible solution is using a 2-dimensional network as implemented in \cite{merolla2014million, davies2018loihi, furber2014spinnaker}. In \cite{furber2014spinnaker}, the 2-dimensional topology is further extended to a 3-dimensional Torus alongside additional diagonal connections and emergency routes. Such implementations involve each spike event to be transmitted as a packet in the mesh. 
 This and similar solutions with more involved networks are subject to  
 congestion in the packet traffic, eventually leading to packet drops. 
 
 Instead, we propose to use the Generalized De Bruijn Graph (GDBG) topology \cite{debruijn} to connect CRAM arrays. GDBG has been used in High Performance Computing and it is shown to be the near-optimal for load-balanced networks \cite{gdbgOptimality}. The first advantage of GDBG is that each vertex has four edges at most, which is comparable to a 2-dimensional mesh. However, the connected vertices are not necessarily in the neighborhood of each 
 other,
 which is defined in a lexicographical ordering sense. The second advantage is that the shortest path between any two vertices is $log_2 N$ for a graph with $N$ vertices. 


Since the connection pattern in GDBG is fixed, we can think about it as
the same repeated pattern allowing each neuron to be connected to all other neurons in $log_2 N$ steps. 
%
{Figure \ref{fig:gdbg2} depicts an example connection scheme for 16 neurons (labeled with numbers 1-16).
in upper-left. 
When the connection scheme is expanded to $\log_2 16=4$ time steps, we obtain the FFT-like diagram in upper-right. Indeed, this connectivity scheme is similar to Singleton's FFT \cite{singleton1967method}, which has a repetitive fixed pattern for each FFT \emph{stage}, i.e., a set of operations with no internal data dependency. The upper-left portion of Figure \ref{fig:gdbg2} shows the resulting hierarchical architecture, where neurons are grouped 
in subsets of 4,
as indicated with colored borders. The bottom portion of Figure \ref{fig:gdbg2} visualizes such an implementation where each edge corresponds to sets of wires between neurons. 
We have the connections between 4-neuron subsets on the bottom-left; and 
each neuron and its connections within the subset, on bottom-right. Applying such a quadtree decomposition, it is possible to map any $N$-neuron network in $\log_4 N$ steps.
}

In our design, each CRAM array, which corresponds to a neuron, is connected to one another in GDBG topology. Each connection consists of $j$ wires, which is equal to the maximum number of presynaptic neurons. Once the computation is done, the routing operations are initiated. The routing consists of $log_2 N$ stages. If the current stage $c\in [1,log_2 N]$ is smaller or equal to the number of presynaptic neurons $j$, then each {array} (i.e., neuron) concatenates the input spike trains and transmits them to the connected arrays for the next stage. This operation involves only reading incoming spike trains and writing them in the predetermined locations. Therefore no extra logic is involved if $c \leq \log_2{j}$. However, if $c>\log_2{j}$,  each array combines the input spike trains using stored address values. Each array therefore reads $\log_2{j}$-bit address values for each spike and copies the spike train to the corresponding addresses. Hence, upon initialization, each array has to store $(\log_2{j})(\log_2{N}-\log_2{j})$ bits of address values. After routing is complete, the next spike computation starts, as discussed in Section \ref{sec:cramsnn:iafinmem}.

\begin{figure}[htp!]
    \centering
    \includegraphics[trim={0cm 2.4cm 13.8cm 0cm},clip,width=7.5cm]{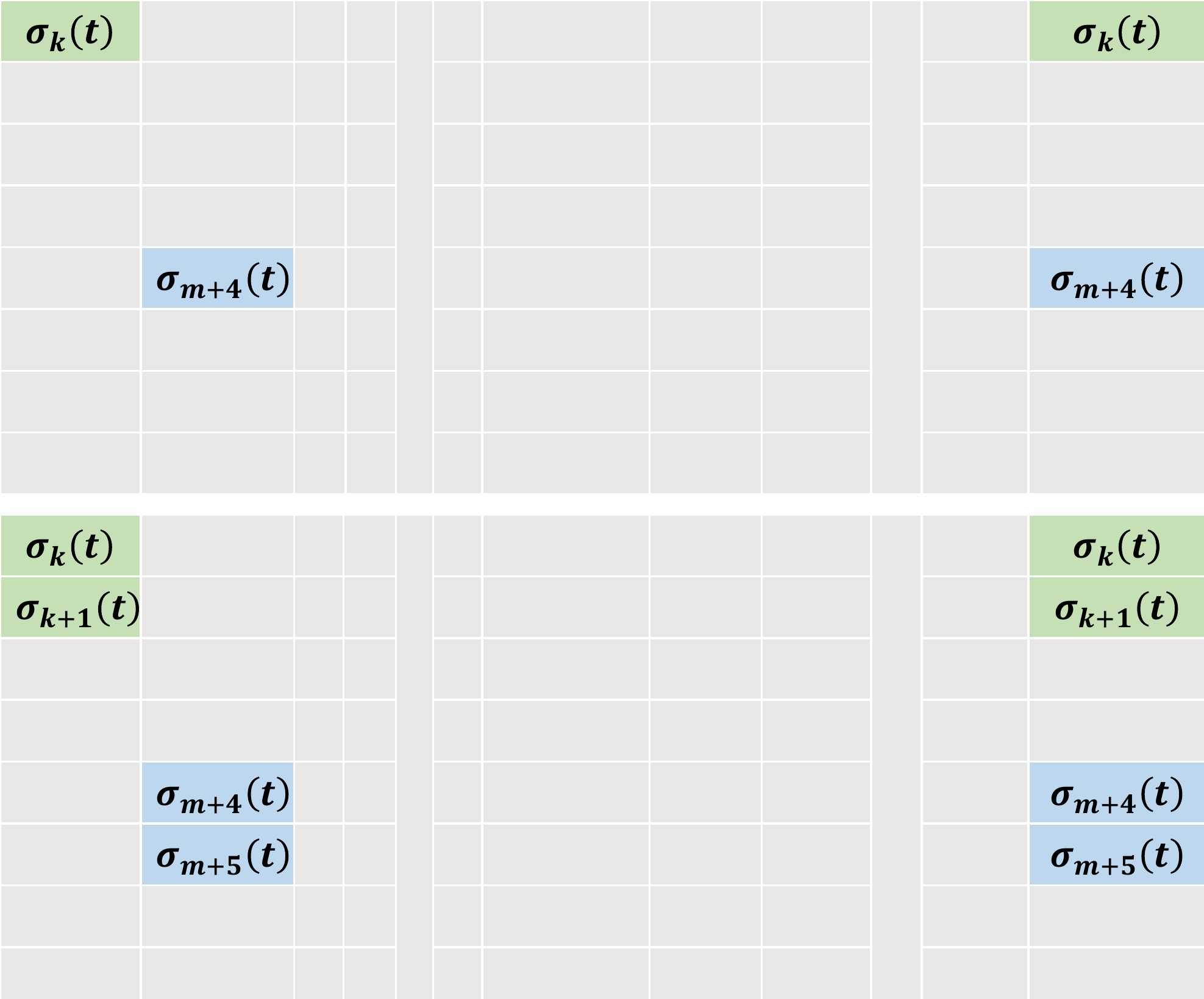}\\
    \vspace{0.03in}
    \includegraphics[trim={0cm 0cm 13.8cm 0cm},clip,width=7.5cm]{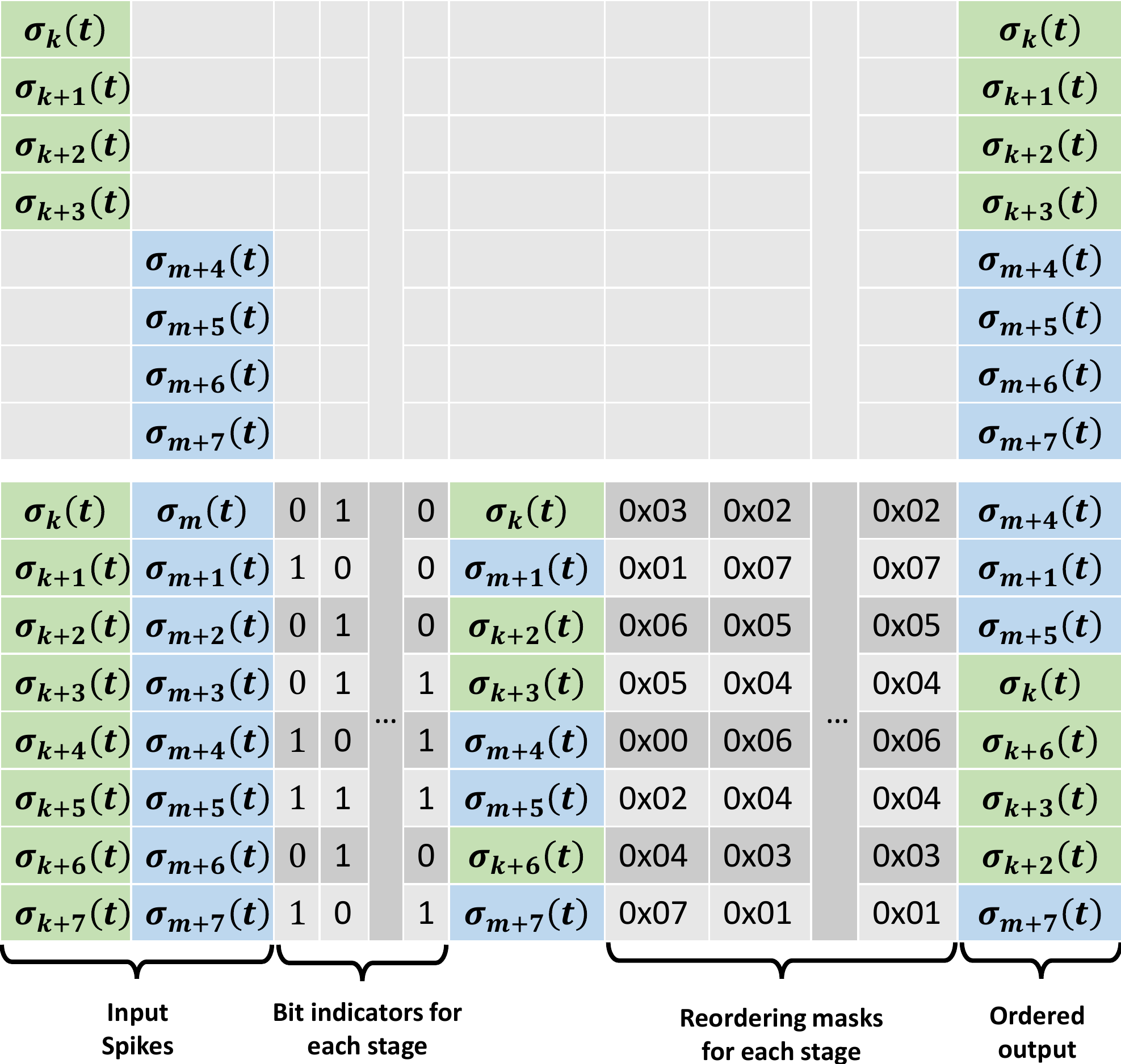}
    \caption{Routing stages for an example 16-neuron (-array) architecture with maximum 8 presynaptic spikes.}
    \label{fig:routing}
\end{figure}

{Fig. \ref{fig:routing} demonstrates 4 routing stages of an example 16-neuron architecture, i.e., $j=8$ and $N=16$. In Stage 1 ($c=1$), the two spikes (highlighted in green and blue) coming from two neurons are combined and transmitted. In Stage 2 ($c=2$), the number of incoming spikes doubles and combined spikes are forwarded to Stage 3. When $c=3$, each incoming spike train has a size of four, and they are forwarded to the next stage. In Stage 4, however, $c=4$ is greater than $\log_2{j}=3$ so half of the incoming spike trains should be discarded and the remaining spikes should be forwarded. For this purpose, we use a bit indicator for each stage, as shown in Figure \ref{fig:routing}.
The bit indicators select one of the input spikes (highlighted in green and blue). The selected input spike train is read and then written to the ordered output in the corresponding address specified in the current stage's reordering (bit)mask. For this example, there is only one set of reordering masks and bit indicators, however, in general there can be $\log_2{N}-\log_2{j}$ different sets of $(\log_2{j})$-bit masks and bit indicator sets.}

{\em Our GDBG based design is efficient since we need $2N$ connections for $N$ neurons instead of $\binom{N}{2}$; we avoid traffic handling which is required to implement NoC based architectures; and we can synchronize the whole system in deterministic time as each routing cycle entails a fixed amount of routing steps, which is $\log_2{N}$ for $N$ neurons.}

\begin{figure*}[ht!]
    \centering
    \includegraphics[trim={16.4cm 15.9cm 29.0cm 0cm},clip,width=15.1cm]{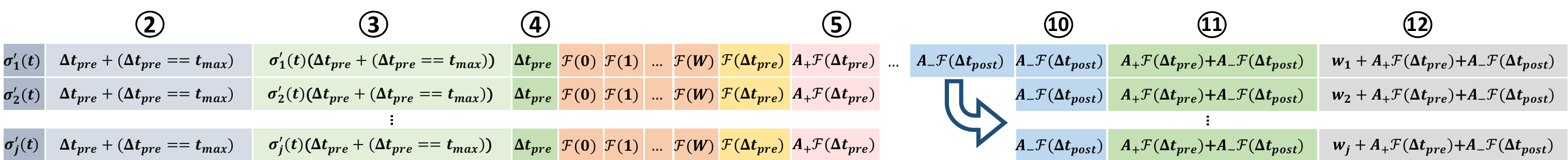}
    \caption{Data layout for STDP engine in CRAM (transposed).}
    \label{fig:stdp3_layout}
\end{figure*}

\begin{figure}[ht!]
    \centering
    \includegraphics[trim={0cm 0cm 57.5cm 0cm},clip,width=6.9cm]{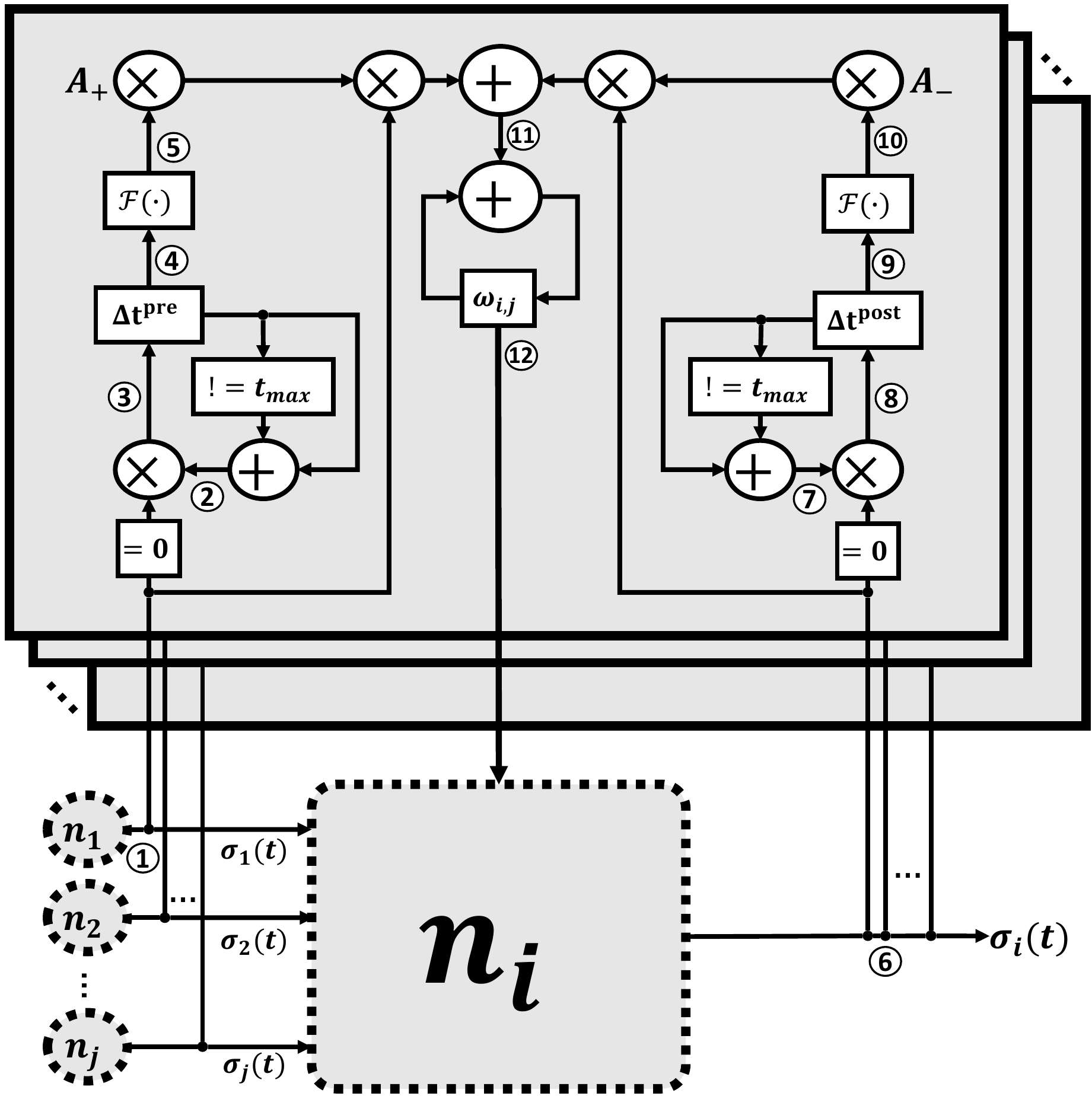}
    \caption{Basic STDP diagram.}
    \label{fig:stdp_alternative}
\end{figure}

\subsection{STDP Learning Engine}

\subsubsection{Initialization}
{
Initialization steps for the STDP engine are similar to 
the feedforward case (Section~\ref{sec:cramsnn:ffinit}). Note that $\mathcal{F(\cdot)}$  is only a scaled version of $\alpha_u(t)$ (Table~\ref{tab:params_defs}). Therefore we re-use the lookup table for $\alpha_u$ and scale any fetched value from it by multiplying the fetched value with $\frac{1}{\tau_u}$. {We also initialize all parameters and constants from Equation~\ref{eq:stdp} such $A_\pm$.}} 

\subsubsection{Data flow}
{In STDP engine, every time spike distribution is completed, presynaptic (postsynaptic) part of the operations is executed, as shown in the left (right) half in Figure \ref{fig:stdp_alternative}, where the input is the spikes and the output is the weight. This effectively performs Equation \eqref{eq:stdp}.}
Note that the data dependency on presynaptic and postsynaptic spikes does not induce additional memory accesses since they are already stored in the same array in our mapping from Section \ref{sec:cramsnn:iafinmem}. 
In the transposed layout, 
\circledwbg{1} corresponds to the presynaptic spikes. \circledwbg{2} is the incremented time difference since the {arrival} 
of the {latest} presynaptic spike and \circledwbg{3} resets the $\Delta t^{pre}$ if the input neuron spikes. Then \circledwbg{5} is obtained by a lookup table implementation of $\mathcal{F(\cdot)}$ function which is fed by \circledwbg{4}. These steps are similar for \circledwbg{7}, \circledwbg{8}, \circledwbg{9}, and \circledwbg{10} except that these calculations are only performed in one row and the output is copied to all rows as shown in the layout in Figure \ref{fig:stdp3_layout}. Finally, $\Delta \omega$ is added to $\omega$ as shown in \circledwbg{11} and \circledwbg{12}. 

\subsubsection{Low level operations}
{Addition and multiplication operations in the STDP are similar to the feedforward case as discused in Section \ref{sec:lif_lowlevel}. Comparators are implemented using a cascade of \texttt{AND} gates.}

\section{Evaluation} \label{sec:eval}

In order to evaluate our design, we perform energy and performance analysis based on the low level elementary operations such as Boolean gate implementations
in the context of the 
proposed CRAM-based SNN architecture.
Configuration parameters for the simulations are given in Table \ref{tab:params}. For experiments, we have four different configurations where STT-M and SHE-M correspond to the current conservative estimates while STT-F and SHE-F reflect near-term future expectations for STT and SHE based MTJs. For array characteristics such as read/write timing, we use 
NVSim \cite{nvsim}. {Each array has an \emph{array controller} which is responsible for  driving logic lines, bitlines, and wordlines.}
We take the overhead resulting from driving such lines into consideration. For synaptic events, we stick to a conservative set-up where  each presynaptic neuron spikes in each time step and each neuron is connected to the maximum possible number of neurons. Since we base our neuron model on Loihi's implementation, we compare our results with the energy and time figures reported in  \cite{davies2018loihi} and \cite{lines2018loihi}, using 
$L=1024$ and $F=1024$ 
for 1-bit synaptic weights (in their design, $L$ is the neuron count per core; $F$, the postsynaptic fanout).

\begin{table}[ht]
    \centering  
    \caption{Configuration Parameters}
    \scalebox{.9}{
    \begin{tabular}{c|c|c}
         \textbf{Parameter} & \textbf{STT-M, SHE-M} & \textbf{STT-F, SHE-F}\\
         \hline
         P state resistance & $3.15~k\Omega$ &  $7.34~k\Omega$ \\
         \hline
         AP state resistance & $7.34~k\Omega$ &  $76.39~k\Omega$ \\
         \hline
         Switching Time & 3 ns & 1 ns \\
         \hline
         Switching Current & $40~\mu A$ & $3~\mu A$ \\
         \hline
         Bulk preset current limit & \multicolumn{2}{c}{30 mA}\\
    \end{tabular}
    }
    \label{tab:params}
\end{table}

\begin{figure}[ht]
    \centering
    \includegraphics[trim={0.1cm 0cm 0.4cm 0.4cm },clip,height=3.2cm]{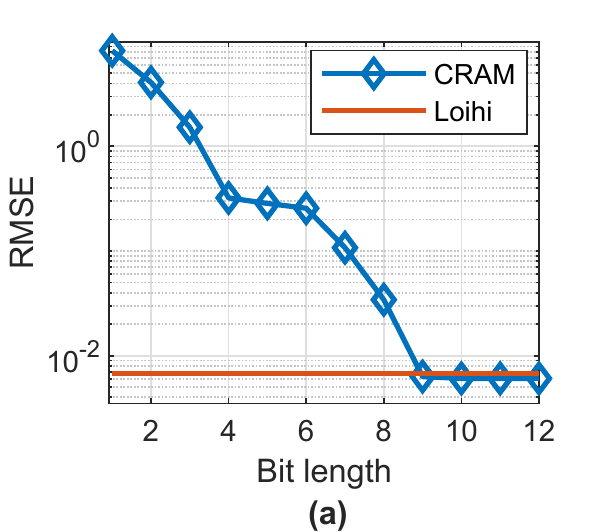}
    \includegraphics[trim={0cm 1.1cm 0.5cm 0.3cm },clip,height=3.2cm]{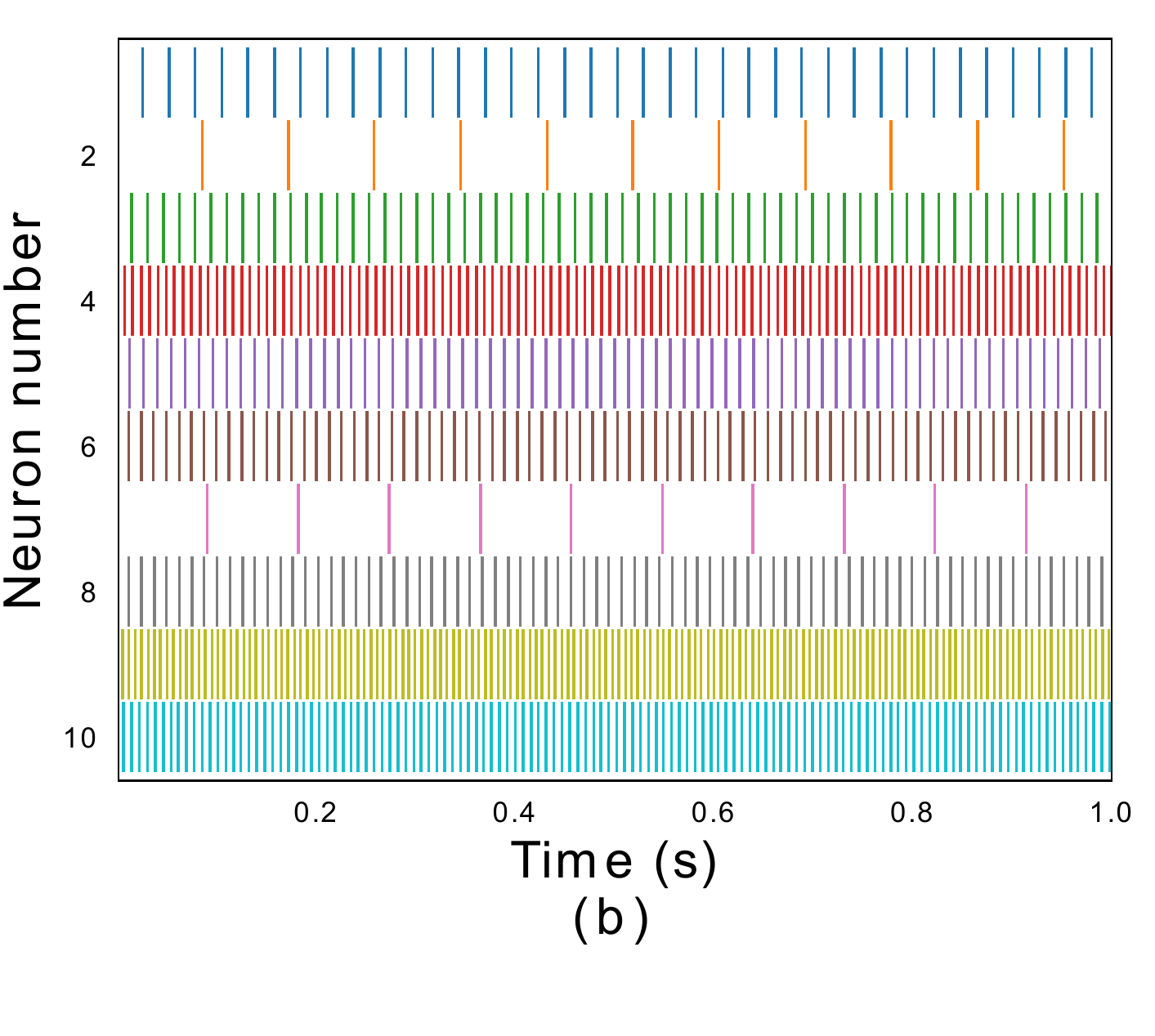}
    \caption{(a) RMSE vs. $\log_2 L_f$ bit length for an example 10-neuron CRAM-SNN compared to 9-bit Loihi. (b) The spike-time diagram.}
    \label{fig:accuracy}
\end{figure}

\begin{figure*}[ht]
    \centering
    \includegraphics[height=6.0cm]{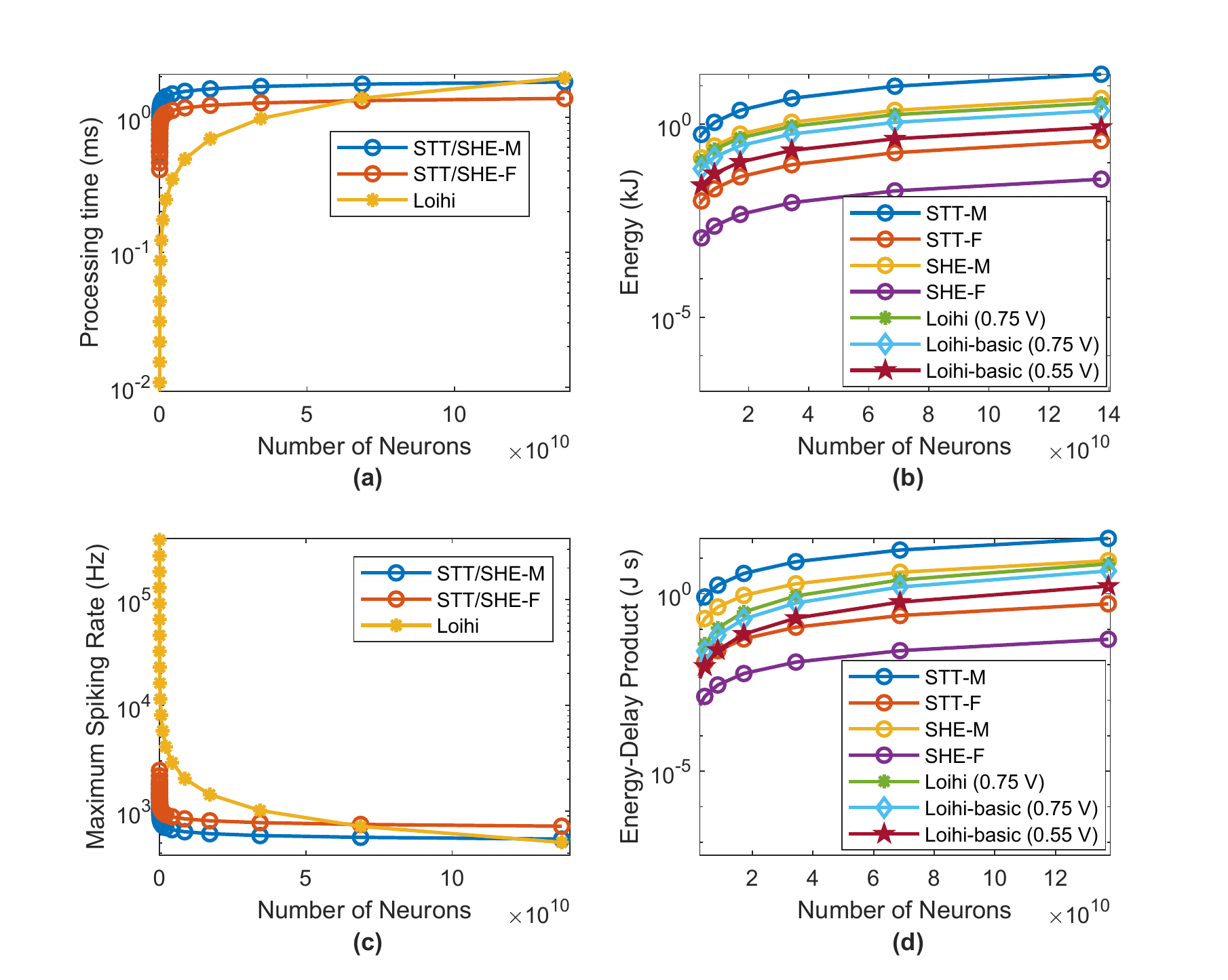}
    \includegraphics[height=6.0cm]{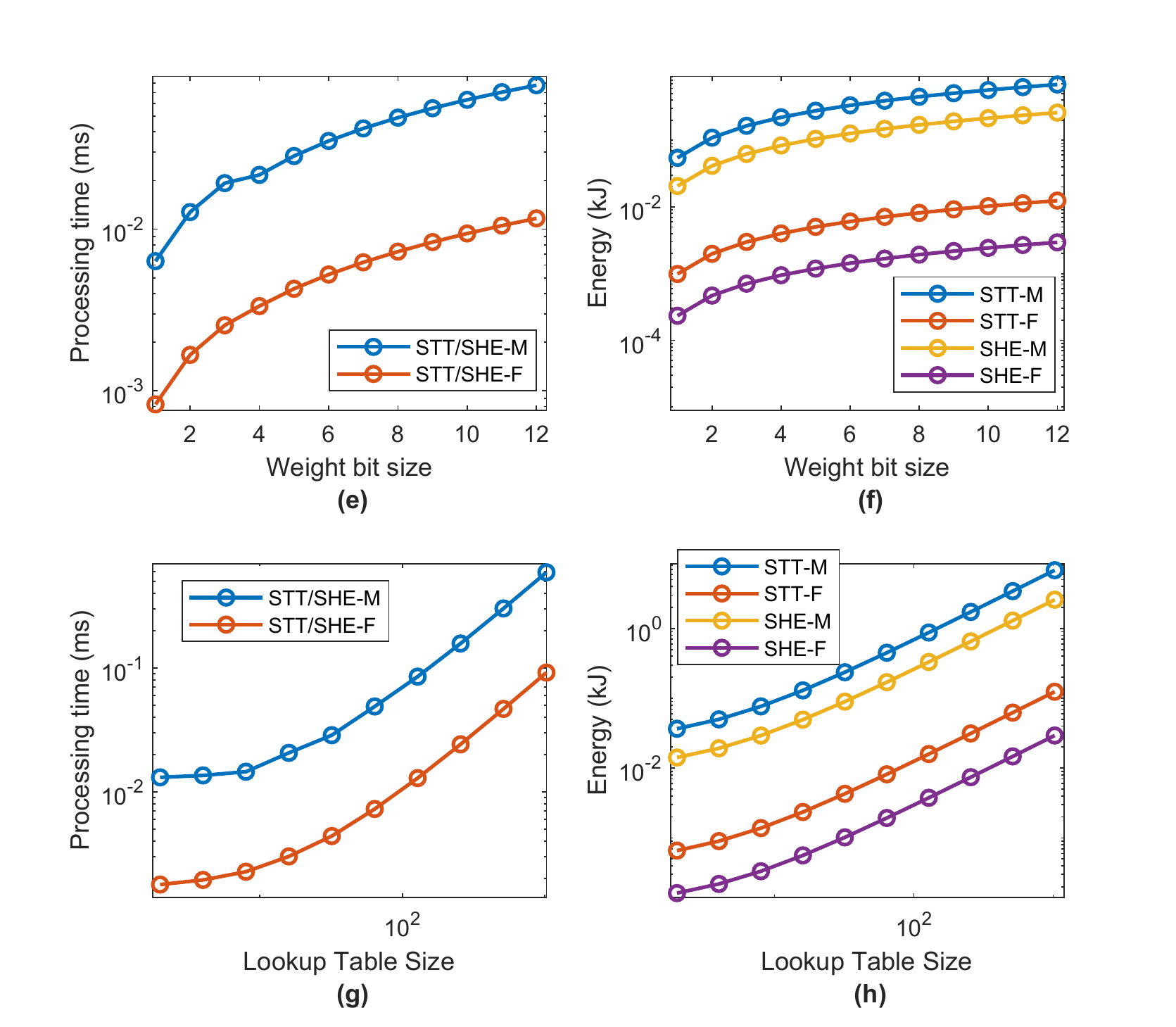}
    \caption{Sensitivity to number of neurons,  weight bit length and $\alpha_u$ table size.}
    \label{fig:plots}
\end{figure*}

For accuracy analysis, we use Loihi library of Nengo framework \cite{bekolay2014nengo} with Loihi configuration. Modifying Nengo Loihi's implementation, we analyze the spiking accuracy for our lookup table based limited precision design. {This analysis is only used for Leaky Integrate-and-Fire feedforward implementation.}
As shown in Figure \ref{fig:accuracy}, RMSE error is comparable to Loihi if at least 9 bits are allocated for each lookup table entry. {Although the accuracy loss is significant in higher bit lengths, Figure \ref{fig:accuracy} (a) shows that a similar bit length to Loihi implementation provides similar accuracy to Loihi.} Figure \ref{fig:accuracy} (b) features 10-neuron spike-time diagram used for the analysis. 

\begin{table}[ht]
    \centering
    \caption{Evaluation Results}
    \scalebox{.93}{
    \begin{tabular}{c|c|c|c|c}
         Metric &  STT-M & STT-F & SHE-M & SHE-F\\
         \hline
         Execution Time ($\mu$s) & 6.342 & 0.825 & 6.342 & 0.825 \\
         Maximum {Spiking Rate} (KHz) & 157.7 & 1212.3 & 157.7 & 1212.3\\
         Energy (J) & 54.51 & 0.989 & 20.61 & 0.234\\
         EDP ($\mu$Js) & 345.72 & 0.816 & 130.74 & 0.193\\
    \end{tabular}
    }
    \label{tab:results}
\end{table}

Table \ref{tab:results} summarizes the results for 1 Billion neurons,  where the maximum number of presynaptic neurons is 1024, bit length is 1 and the filter (lookup table) size is 64. Each CRAM array has 1024$\times$ 512 cells.

{It can be seen that although all CRAM configurations can provide enough performance to implement SNN operations within a biologically plausible time budget, i.e., a spiking rate of several kHz, SHE-F configuration provides the lowest energy consumption and fastest operation.}

\begin{figure} [ht]
    \centering
    \includegraphics[trim={1.5cm 0cm 1.5cm 0cm },clip,width=8.8cm]{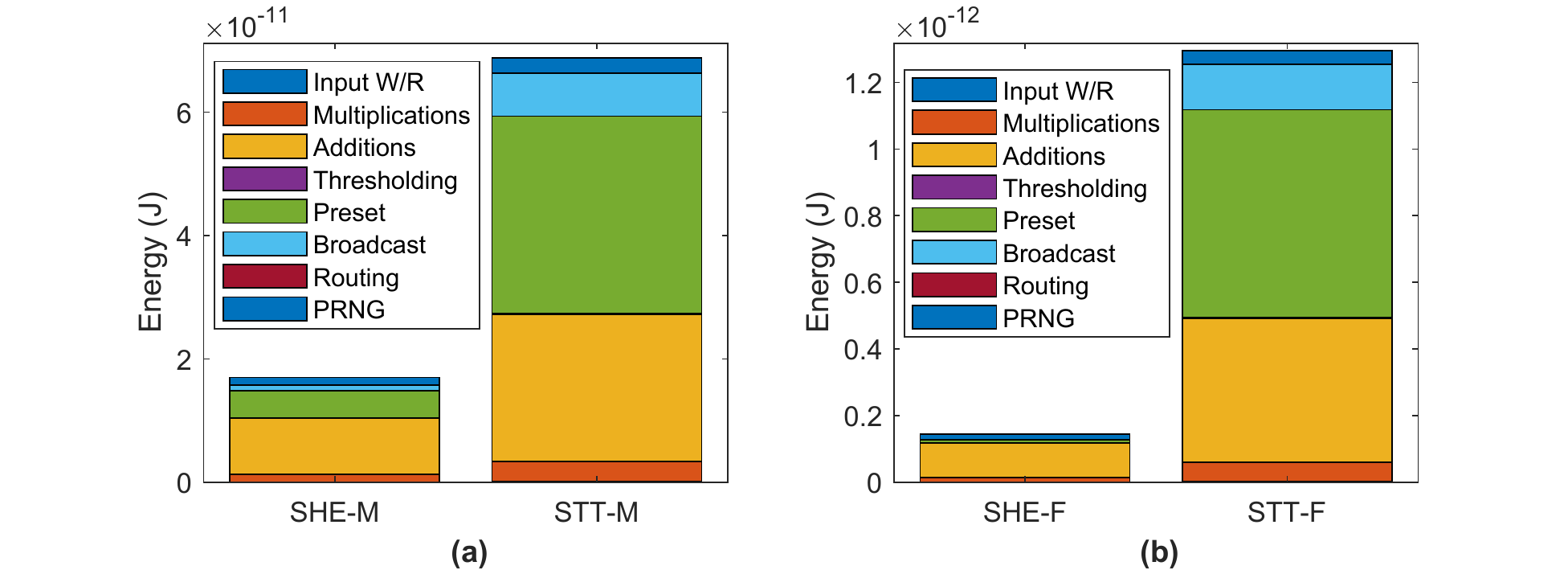}
    \caption{Overall energy breakdown of operations in a single instance of spiking events.}
    \label{fig:breakdown_power}
\end{figure}

Figure \ref{fig:breakdown_power} shows the energy breakdown for a single spike operation with 1 bit weights. We observe that preset and additions dominate the overall energy followed by multiplication (i.e., \texttt{AND}) operations, pointing to further hardware optimization opportunities. 

Figure \ref{fig:plots} shows execution time, maximum {spiking rate},
energy and energy-delay-product (EDP) for the case where maximum number of presynaptic neurons is 1024 and the bit length is 8. 
{Figure \ref{fig:plots}(a) and \ref{fig:plots}(c) are complementary figures for -M and -F variants compared to Loihi model for similar parameters. Although Loihi is faster for larger number of neurons because of its faster computation, CRAM variants surpass Loihi when neuron count approaches to hundred billions thanks to our routing architecture. Figure \ref{fig:plots}(b) and \ref{fig:plots}(d) capture energy and EDP's sensitivity to neuron counts in several Loihi models with different voltages. For very large neuron counts, the best performing CRAM variants have lower energy consumption than Loihi, and higher energy efficiency.}  We also sweep bit length of weights and table size for $\alpha_u$, as shown in Figure \ref{fig:plots}(e)--(h). Figures {\ref{fig:plots}(e) and \ref{fig:plots}(f) show processing time and energy consumption for different weight bit lengths; and Figures \ref{fig:plots}(g) and \ref{fig:plots}(h) capture the processing time and energy for various lookup table sizes in the feedforward design.} In all cases, {STT/SHE-F or }SHE-F configuration provide the {lowest processing time and energy when  compared to other CRAM variants, emphasizing the effect of the cell technology.} 

Overall, when compared to the best performing 
Loihi baseline (0.55V) with {a  fanout of 1024}, 
and 1-bit synaptic weights, 
SHE-F configuration consumes  $26.13\times$ less energy. 
At the same time, SHE-F is $3.99\times$ more energy efficient (in terms of EDP, energy-delay-product) {for the feedforward implementation as captured in Figure \ref{fig:plots}}. 

\begin{table}[ht!]
    \caption{Energy, latency, and EDP for single spike operation.}
    \scalebox{.9}{
    \begin{tabular}{c|c|c}
         \textbf{Synaptic Parameter} & \textbf{SHE-F} & \textbf{Loihi}\\
         \hline
         Spike Operation Energy & 143.81 fJ & 23.6 pJ\\
         Spike Operation Time & 499.86 ns & 3.5 ns\\
         Spike Operation EDP &7.19e-20 Js & 8.26 e-20 Js\\
         STDP Update Energy & 0.33 pJ & 120 pJ\\
         STDP Update Time & 1321.7 ns & 6.1 ns\\
         STDP Update EDP & 4.38e-19 Js & 7.32 e-19 Js
    \end{tabular}
    }
    \label{tab:perf_params}
\end{table}

Table \ref{tab:perf_params} summarizes low level performance parameters of our best case CRAM-SNN design for 1-bit weights where $L_f=32$ for feedforward computations. Also tabulated are  Loihi's corresponding low-level performance parameters for comparison. STDP operations are performed at 10-bit precision. 
{$L_f=32$ is chosen to capture enough accuracy to model $\alpha_u(t)$ while lowering logic complexity to maintain a low energy consumption. 10-bit precision suffices to provide enough bandwidth for updating up to 9 bit weights, which is meaningful for a fair comparison.} 
Although being slower, our design consumes significantly less energy, which leads to a lower EDP for a single spike operation with a single presynaptic connection for 1 bit weights. 

\begin{figure}[ht!]
    \centering
    \includegraphics[trim={1.3cm 0cm 1.5cm 0cm },clip,width=8.6cm]{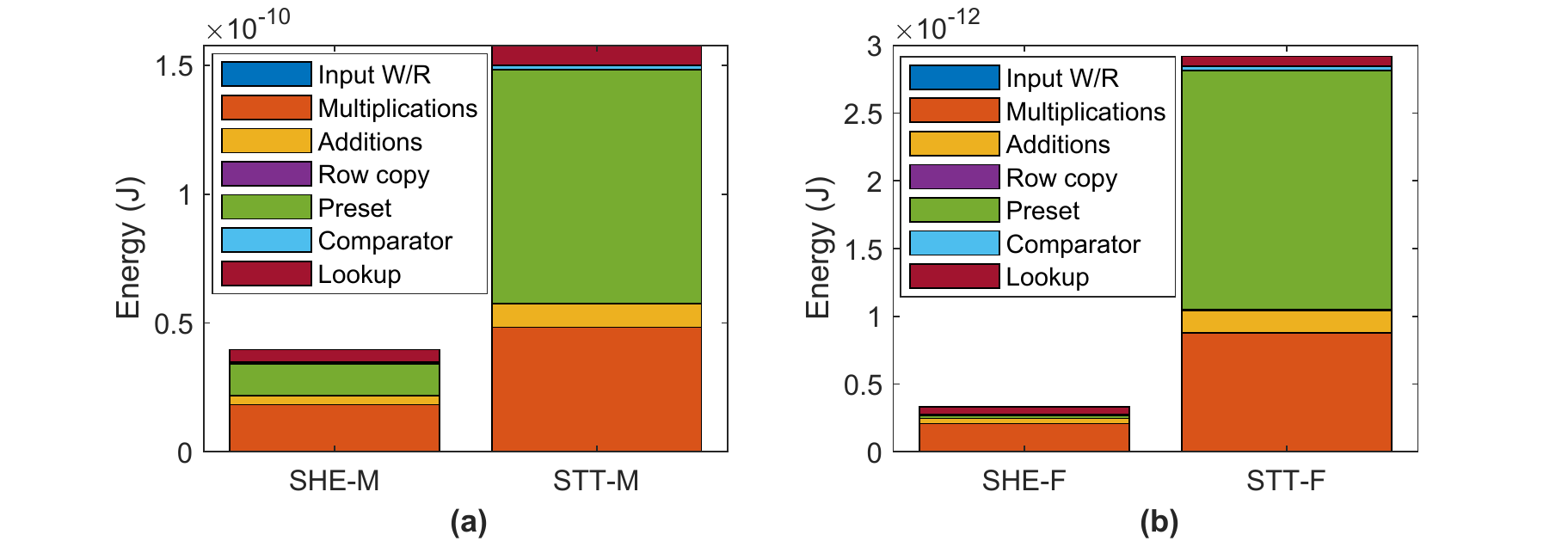}
    \caption{Single-spike energy breakdown of STDP weight update operation.}
    \label{fig:stdp_energy_breakdown}
\end{figure}

\begin{figure}[ht!]
    \centering
    \includegraphics[trim={0.5cm 0cm 1.1cm 0cm },clip,width=8.6cm]{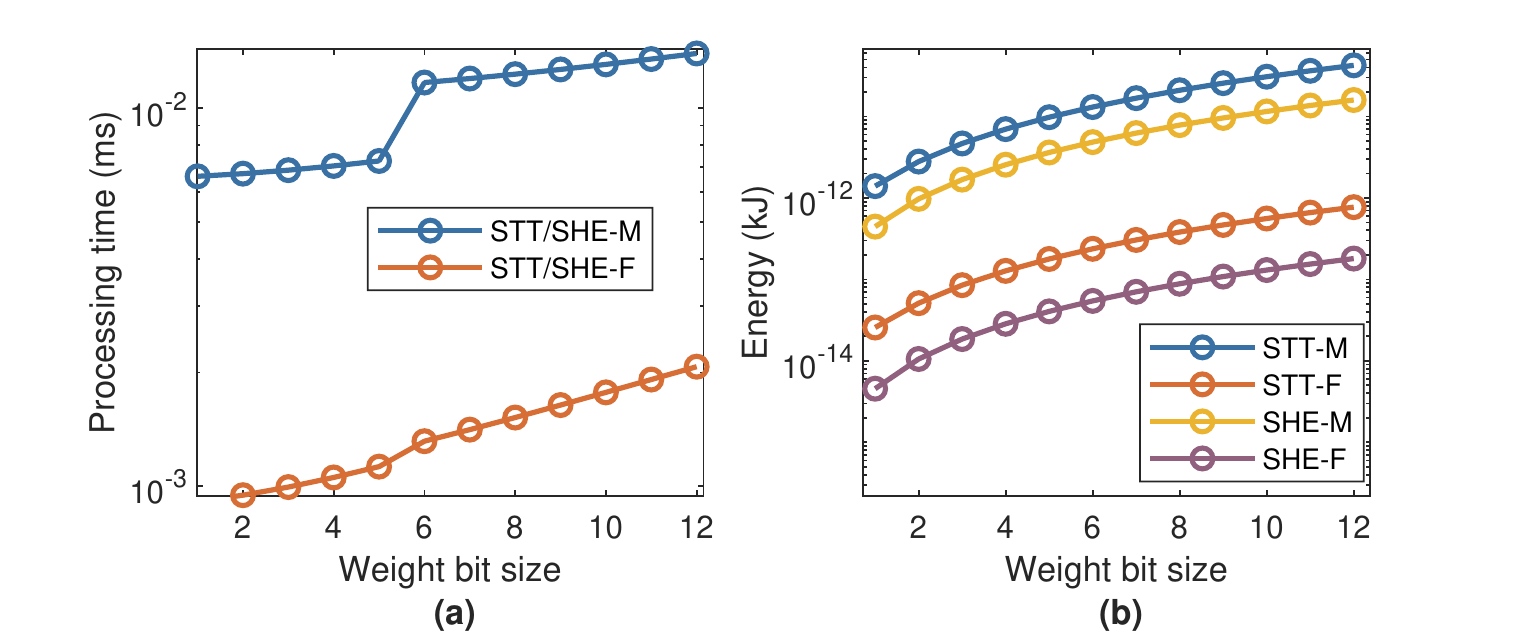}
    \caption{(a) Processing time vs. weight bit length. (b) STDP update energy vs. weight bit length.}
    \label{fig:stdp_w_sweep}
\end{figure}

Figures \ref{fig:stdp_w_sweep}(a) and \ref{fig:stdp_w_sweep}(b) show the STDP processing time and energy for larger bit lengths.
{Processing times in \ref{fig:stdp_w_sweep}(a) irregularly increase around 6-bits as CRAM preset operations become overwhelming,
however, this does not effect the energy consumption pattern in \ref{fig:stdp_w_sweep}(b) where the SHE-F configuration outperforms the other CRAM variants.} Energy breakdown of STDP operations are given in Figures \ref{fig:stdp_energy_breakdown}(a) and \ref{fig:stdp_energy_breakdown}(b). Similar to the feedforward case, energy consumption is mostly dominated by preset, addition and multiplication operations in STDP weight update. 
{Low level performance results further show that 
our gains are not
only due to the proposed connectivity scheme, but also because of the low energy, massively parallel logic operations CRAM enables.}

\section{Conclusion} \label{sec:conclusion}
We introduce a CRAM-based SNN architecture, where each neuron is processed inside a single CRAM {array} while the {arrays} are connected to each other in a GDBG 
topology. 
{
We thereby achieve a limited full connectivity by using only
$2N$ connections in a $N$ neuron network,  instead of $\binom{N}{2}$, while exploiting the massive intra- and inter-array parallelism and energy efficiency of CRAM for logic operations.} 
Our best configuration results in $164.1\times$ less energy consumption and {similar} EDP {for feedforward operations; and $361.77\times$ less energy consumption and $1.66\times$ lower EDP for learning (via STDP) when compared to the alternative Loihi architecture}. 


\bibliographystyle{ACM-Reference-Format}
\bibliography{bibliography}

\end{document}